\documentclass[aps,pra,preprint,groupedaddress,showpacs,%
tightenlines,
nofootinbib,floatfix]{revtex4}

\usepackage{upgreek}  
\usepackage{graphicx}
\usepackage{amsmath,amssymb,amsfonts,bm}

\begin{document}

\title{Influence of epithermal muonic molecule formation on kinetics of
  the $\boldsymbol{\mu}$CF processes in deuterium}

\author{A.~Adamczak}
\email{andrzej.adamczak@ifj.edu.pl}
\affiliation{Institute of Nuclear Physics, Polish Academy of Sciences,
  Radzikowskiego 152, PL-31342~Krak\'ow, Poland}

\author{Mark P. Faifman}
\email{mark@rogova.ru}
\affiliation{Research Coordinative Center "MUCATEX", 
  Moscow, 123098, Russia}

\date{\today}

\begin{abstract}
  The non-resonant formation of $dd\mu$ molecules in the loosely bound
  state in collisions of non-thermalized $d\mu$ atoms with deuterium
  molecules D$_2$ has been considered. The process of such a~type is
  possible only for collision energies exceeded the ionization potential
  of D$_2$. The calculated rates of $dd\mu$ formation in the
  above-threshold energy region are about one order of magnitude higher
  than obtained earlier.

  The role of epithermal non-resonant $\mu$-molecule formation for the
  kinetics of $\mu$CF processes in D$_2$ gas was studied. It was shown
  that the non-resonant $dd\mu$ formation by $d\mu$ atoms accelerated
  during the cascade can be directly observed in the neutron time
  spectra at very short initial times.
\end{abstract}

\maketitle

\section{Introduction}
\label{intro}

The studies of various reactions with negative $\mu^{-}$ muons in
a~deuterium target and the muon catalyzed fusion ($\mu$CF) phenomenon in
particular give rise to special interest in the
$dd\mu$-molecule-formation processes (see works~\cite{ZG,Bal} and
references therein).  In collisions of the $d\mu$ atoms with the
deuterium D$_2$ molecules, the $dd\mu$ molecules are formed in one of
the five bound states, which are defined by the different
rotational~($J$) and vibrational~($\upsilon$) quantum
numbers~\cite{Vin}.
The loosely bound state with binding energy
$|\varepsilon_{J=1,\upsilon=1}|=1.975$~eV refers to $dd\mu$ formation by
the resonant reaction:
\begin{equation}
  d\mu + \mathrm{D}_2\rightarrow[(dd\mu)_{11}\, dee]_{K\nu}^{*} \,,
  \label{e1}
\end{equation}
where the released energy $\varepsilon\approx{}2$~eV is transferred to
the excitation of ro-vibronic~($K\nu$) states of the molecular complex
$[(dd\mu)dee]$, according to the resonance mechanism~\cite{Ves}. At the
temperature conditions of the majority of previous experiments, the
processes~(\ref{e1}) mainly occur for thermalized $d\mu$ atoms in the
ground state. The rate~$\lambda_{dd\mu}$ of resonance
reaction~(\ref{e1}) depends on the target temperature $T$ and is on the
order of $10^6$~s$^{-1}$~\cite{Bal,Men87} for room temperature
$T=300$~K.

In any other ($J\upsilon$) state, the $dd\mu$ molecules are formed via the
non-resonant process~(\ref{e2}):
\begin{equation}
  d\mu+\mathrm{D}_2 \to [(dd\mu)_{J\upsilon}\, de]^{+}+e^{-}\,,
  \label{e2}
\end{equation}
with conversion of the released energy into electron ionization of the
D$_2$ molecule (see references in reviews~\cite{ZG}). The rates of
transitions to all existing $dd\mu$ ro-vibronic states have been
calculated in~\cite{MPF89}. It has been shown that collisions of the
thermalized $d\mu$ atoms with the D$_2$ molecules lead to non-resonant
formation of $dd\mu$ molecules with the rates
$\lambda_{dd\mu}\sim10^4$~s$^{-1}$, whereas these rates are on the order
of $10^6$~s$^{-1}$ for non-thermalized $d\mu$'s.

It should be noted that for thermalized $d\mu$ atoms the non-resonant
$dd\mu$ formation~(\ref{e1}) in the state $J=\upsilon=1$ is impossible.
However, as presented in this work, such a~process is realized for the
non-thermalized $d\mu$ atoms. Also, it is shown that the calculated
non-resonant rates are much higher than the data obtained
in~Ref.~\cite{MPF89}.

\section{Non-resonant formation of $\boldsymbol{dd\mu}$ molecule}
\label{sec:1}


For thermalized $d\mu$ atoms, collision energies $\varepsilon$ are
usually much less than the ionization potential $I_e=15.46$~eV of the
D$_2$ molecule ($\varepsilon\ll{}I_e$). Then the non-resonant $dd\mu$
formation~(\ref{e2}) in the loosely bound state with electron conversion
is impossible ($\varepsilon_{11}<I_e$), and only resonant
formation~(\ref{e1}) is realized. However, the reactions~(\ref{e2}) take
place for $dd\mu$ formation in deeper bound states with binding energies
$\varepsilon_{J\upsilon}\geq I_e$. Besides, when the non-thermalized
$d\mu$ atoms have quite high energies $\varepsilon\geq{}I_e$, the
non-resonant formation of $dd\mu$ molecules in the loosely bound state
($J=1, \upsilon=1$) also becomes possible.

\begin{figure}[htb]
  \includegraphics[width=8cm]{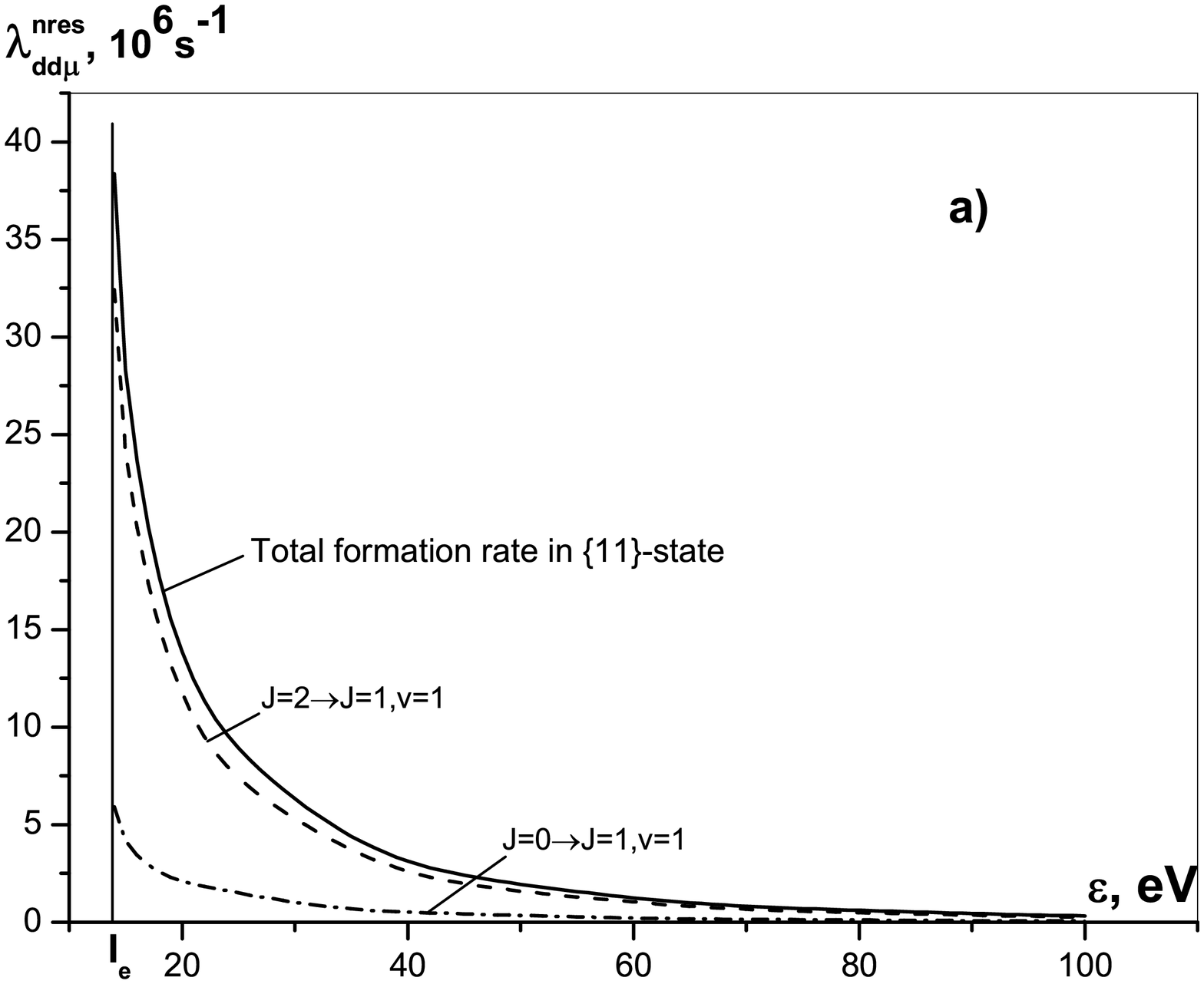}
  \includegraphics[width=8cm]{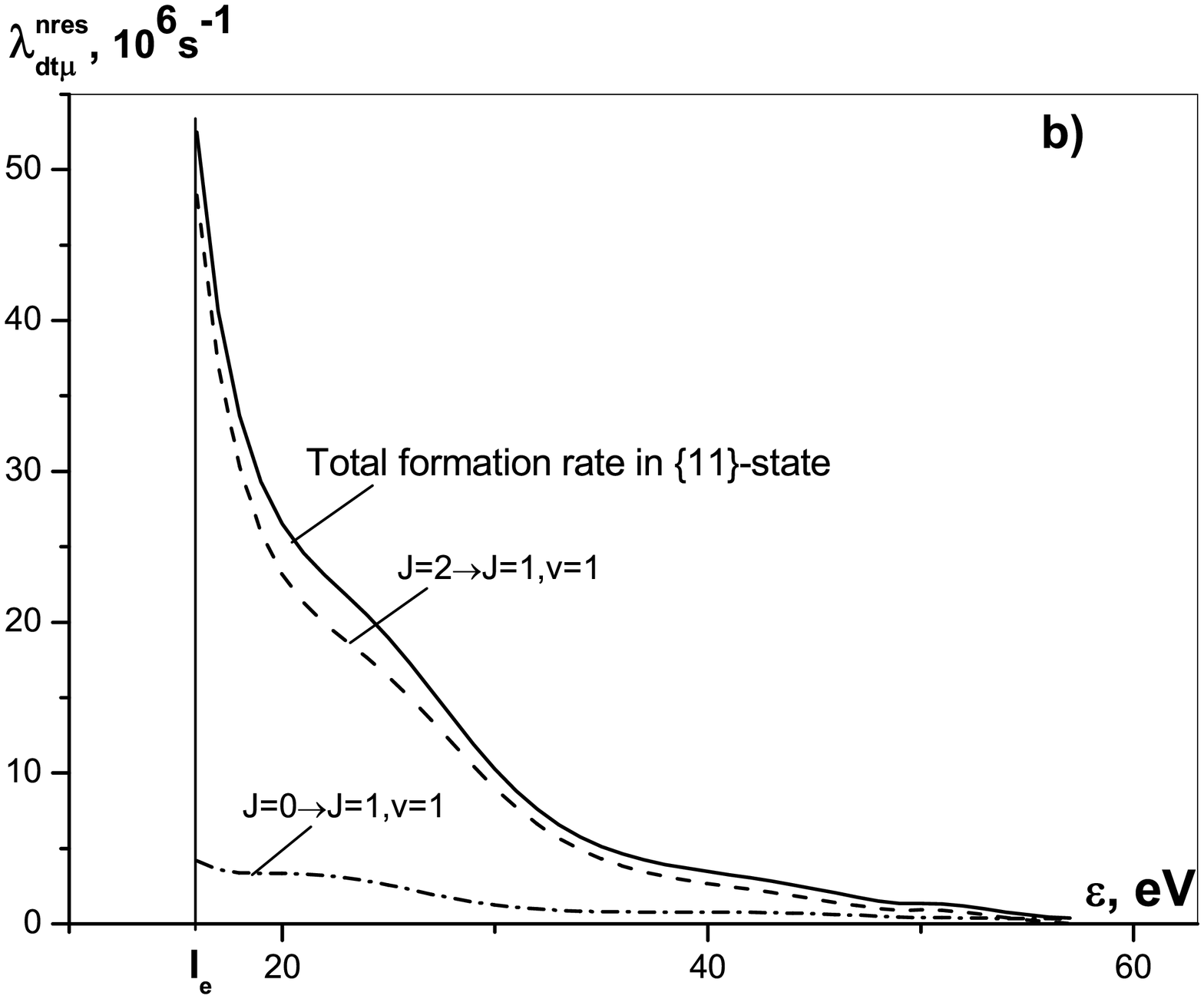}
  \caption{Non-resonant formation rates of $dd\mu$~(a) and $dt\mu$~(b)
    in the state ($J$=$\upsilon$=1). The dash-dotted and dashed lines denote
    the $E1$-transitions from the initial states $J=0$ and $J=2$,
    respectively.}
  \label{fig:1}
\end{figure}
A~method for calculating the non-resonant formation rates in collisions
of the epithermal $d\mu$ atoms with the D$_2$ molecules is analogous to
the method developed in~Ref.~\cite{MPF89}. There it was shown that the
dominating transitions from the scattering states of the $d\mu+d$ system
to the bound states of muonic molecules, with the total orbital angular
momenta $J=1$, are the electric $E1$ transitions only. Such transitions
have been considered in this work.
The corresponding rates are presented in Fig.~\ref{fig:1}a as functions
of collision energy $\varepsilon$ in the center of mass of the system
$d\mu+$D$_2$ (additionally, the analogous dependencies of non-resonant
formation rates of the $dt\mu$ molecules are shown in the
Fig.~\ref{fig:1}b). At $\varepsilon=I_e-|\varepsilon_{11}|$, the plotted
rates have a~typical threshold peculiarity and maximum values, because
of the existence of the loosely bound state~(11).

\begin{figure}[htb]
  \centering
  \includegraphics[width=9.5cm]{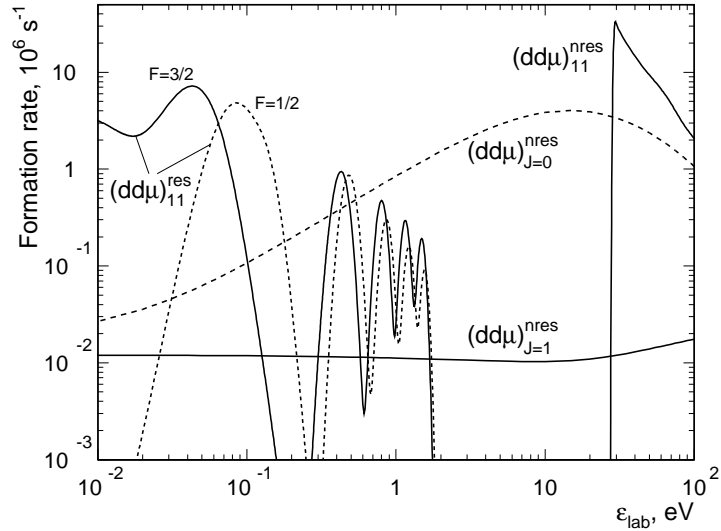}
  \caption{Resonant and non-resonant $dd\mu$ formation rates versus
    $d\mu$ kinetic energy in laboratory system.}
  \label{fig:2}
\end{figure}

In Fig.~\ref{fig:2}, the obtained results are compared with the earlier
calculated total rates of non-resonant $dd\mu$ formation~\cite{MPF89} in
the rotational states $J=1$ and $J=0$, as well as resonant formation in
the ($J=1,\upsilon=1$) state~\cite{AF_09}.
It is apparent that in the above-threshold energy region the calculated
rates of $dd\mu$ formation in the loosely bound (11) state are about one
order of magnitude higher than the rates~\cite{MPF89} of $dd\mu$
formation in the lower state $J=0$. Also, they are more than three
orders of magnitude higher for formation in the $J=1~(\upsilon\neq 1)$
state.  This follows from the fact that the overlap of the wave
functions of the initial and final states of the $d\mu+$D$_2$ system is
much stronger in the case of non-resonant $dd\mu$ formation in the
loosely bound ($J=1, \upsilon=1$) state than in the state ($J=1,
\upsilon\neq 1$).

\section{Demonstration of the epithermal $\boldsymbol{dd\mu}$ 
         formation effect}
\label{sec:2}

\begin{figure}[htb]
  \includegraphics[width=8cm]{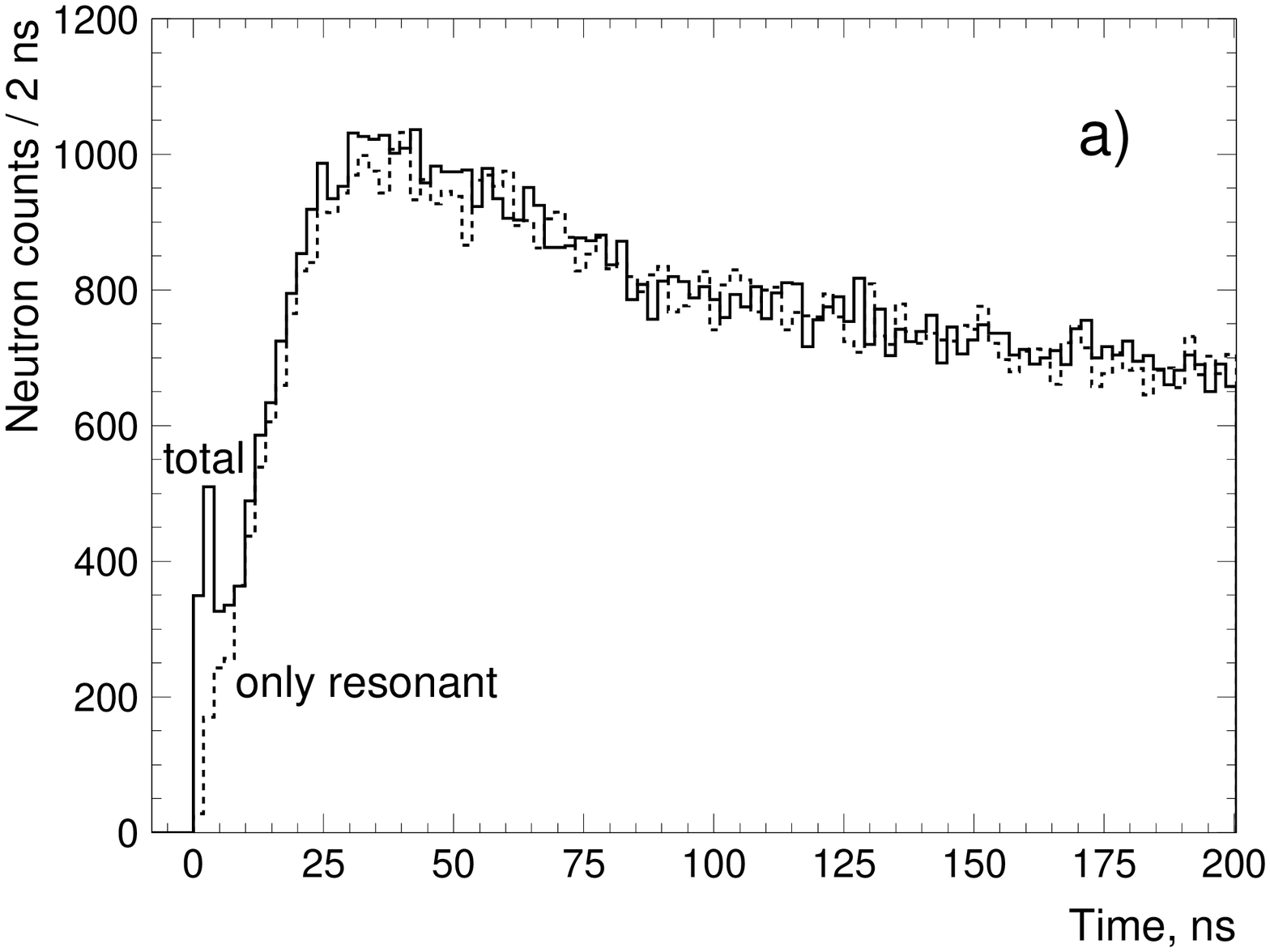}
  \includegraphics[width=8cm]{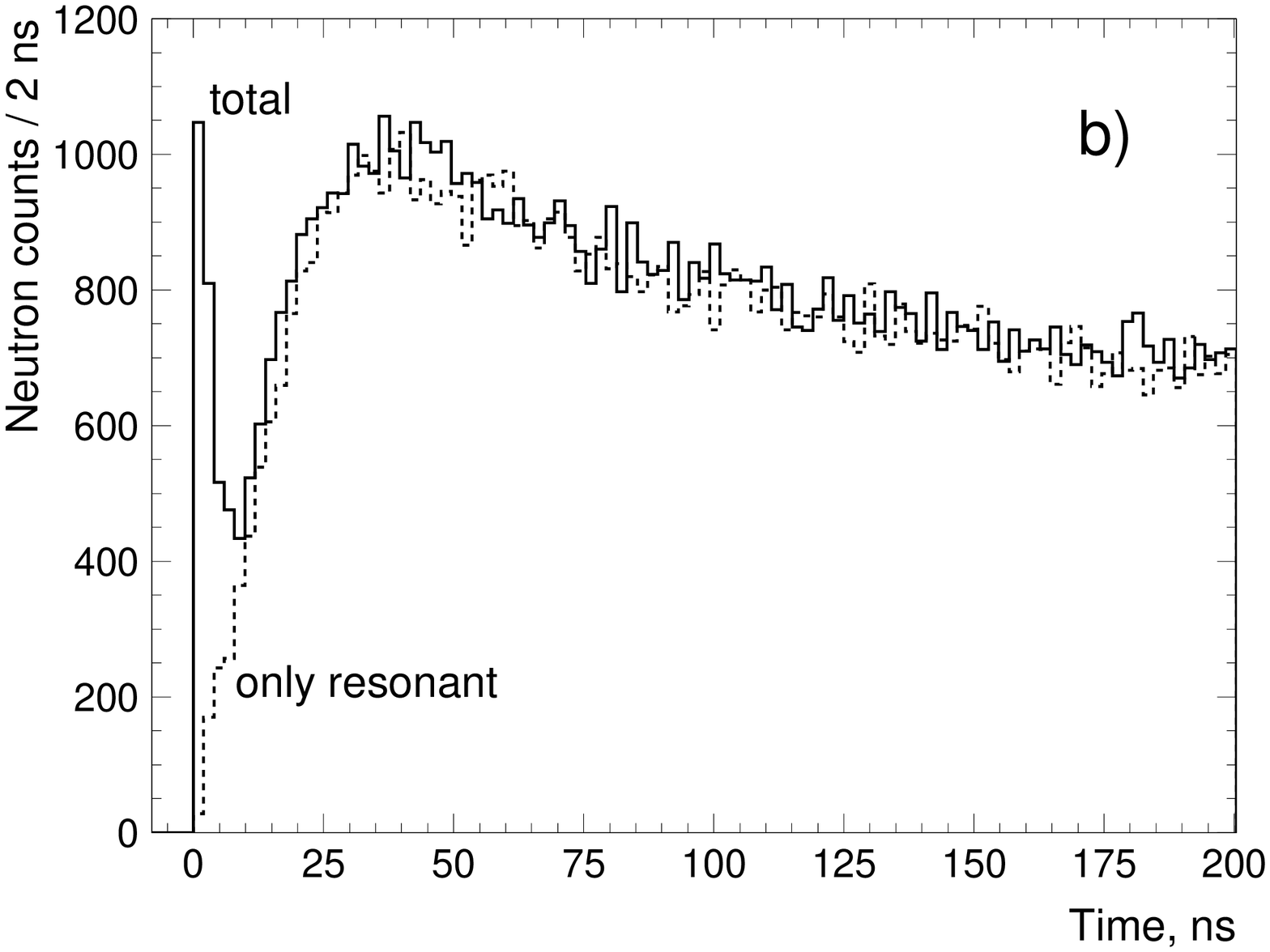}
  \caption{Neutron spectra from $dd$~fusion in D$_2$ at
    $T=40$~K and $\phi=0.05$ (solid line) and the analogous spectrum
    obtained assuming absence of nonresonant $dd\mu$ formation
    (dashed line): a)~the initial mean energy of non-thermalized $d\mu$
    atoms ~$\varepsilon_\text{avg}=10$~eV,
    b)~$\varepsilon_\text{avg}=50$~eV.}
  \label{fig:3}
\end{figure}

The kinetics of $\mu$CF processes in a~pure D$_2$ gas has been studied
in order to take into account effects of the presence of non-thermalized
$d\mu$ atoms. For this purpose, the kinetic-energy distributions of
$d\mu$ atoms in different atomic states, which are established just
after cascade de-excitations of the formed $d\mu$'s, have been
calculated using method~\cite{MF07}. These calculations confirmed that
most of the $d\mu$ atoms in the final 1$S$~state are not thermalized,
due to collisions in the cascade process~\cite{ab_po,faif_jens}. For
simplicity of further numerous calculations, an assumption of a~simple
two-Maxwell shape of the initial energy distribution of $d\mu$
atoms~\cite{AF_09,ab_po} has been employed. One of the Maxwell
components of this distribution corresponds to the non-thermalized
atoms, while the second component describes the thermalized atoms.  The
time spectra of neutrons from the $dd$ fusion in $dd\mu$ have been
calculated by means of Monte-Carlo simulations of the kinetics of
$\mu$CF processes~\cite{AF_09}. They are shown in Figs.~\ref{fig:3}a,b
for the D$_2$-gas target at temperature $T$=40~K and density $\phi$=0.05
(in the liquid-hydrogen-density units). Since the accuracy of
calculating the kinetic-energy distribution of 1$S$ $d\mu$ atoms is
still insufficient, the two average
energies~$\varepsilon_\text{avg}=10$~eV (Fig.~\ref{fig:3}a) and
~$\varepsilon_\text{avg}=50$~eV (Fig.~\ref{fig:3}b) of the
non-thermalized Maxwell component~\cite{AF_09} have been chosen. The
neutron spectra with both the resonant and non-resonant $dd\mu$
formation taken into account are represented by the solid lines, while
the dashed lines have been calculated without the presence of the
non-resonant formation processes.

The~significant effect of the epithermal non-resonant $dd\mu$ formation
is displayed in Fig.~\ref{fig:3} as a~peak at the time $t\approx{}5$~ns.
According to Fig.~\ref{fig:2}, it is clear that such an effect can be
revealed only at short times ($t\lesssim{}20$~ns), when most of the
$d\mu$ atoms are not yet slowed down to the energy region corresponding
to the high peaks of resonant formation. In Fig.~\ref{fig:3}b, the
prompt peak is more pronounced since the non-thermalized fraction of
initial $d\mu$'s corresponds to a~higher mean energy
$\varepsilon_\text{avg}=50$~eV.

The enhancement of the neutron yield from $dd$ fusion at short times,
which was already observed in $\mu$CF experiments in D$_2$ gas (Fig.~11
in Ref.~\cite{Bal}) and was even more pronounced in HD gas (Fig.~18 in
Ref.~\cite{Bal}), confirms the nature of phenomenon considered above.
A~further consistent comparison between the measured and calculated
neutron time spectra for the $\mu$CF in D$_2$ target would enable
drawing a~final conclusion about the mean energy of $d\mu$ atoms in the
$1S$ state, which are accelerated in the cascade processes.

\section{Conclusions}

A study of the kinetics of $\mu$CF processes in D$_2$ gas, in particular
the neutron spectra at short times, revealed the~significant role of the
non-resonant formation of $dd\mu$ molecules at kinetic energies
characteristic to the non-thermalized ground-state $d\mu$ atoms, which
are accelerated during the atomic cascade.

Besides the well-known processes~(\ref{e1}) and~(\ref{e2}) of
$\mu$-molecule formation, a~new possibility of non-resonant formation of
the $dd\mu$ molecule in the loosely bound state in the presence of
non-thermalized $d\mu$-atoms has been considered. The calculated rates
of such formation reach the magnitude
$\lambda_{dd\mu}\sim{}10^6$~s$^{-1}$, near the energy threshold of
reaction~(\ref{e2}). Therefore, these reactions should also be taken
into account in analyses of $\mu$CF kinetics in deuterium, in
particular, at low target densities. Moreover, detailed information on
the experimental short-time neutron spectra would allow to extract the
mean energy of $d\mu$ atoms in the ground state and to estimate the
reliability of various cascade-characteristics calculations.

\section*{Acknowledgments}
  The authors are grateful to Profs. L.I.~Ponomarev and A.A.~Vorobyov
  for helpful discussions. It is a pleasure to thank Drs. M. Jeitler and
  N.I. Voropaev for their keen interest.

%
\end{document}